\let\Xdocument\document
\documentclass{iau}
\let\document\Xdocument

\usepackage{amsmath}
\usepackage{graphicx}
\usepackage{multirow}

\lefttitle{Jain and Tripathy}
\righttitle{Subsurface Flows in Active Regions with Peculiar Magnetic Configurations}

\jnlPage{1}{6}
\jnlDoiYr{2024}
\doival{10.1017/xxxxx}

\aopheadtitle{Proceedings IAU Symposium}
\editors{A. Getling \& L. Kitchatinov, eds.}

\title{Subsurface Flows in Active Regions with Peculiar Magnetic Configurations}

\author{Kiran Jain and Sushanta C. Tripathy}
\affiliation{National Solar Observatory, 3665 Discovery Dr., Boulder, CO 80303, USA}

\begin{document}
\begin{abstract}
 We present analysis of the evolution of subsurface flows in and around active regions 
with peculiar magnetic configurations and compare their characteristics with the 
normal active regions.  We also study the zonal and meridional components of subsurface 
flows separately in different polarity regions separately to better understand their 
role in flux migration. We use the techniques of local correlation tracking and  
ring diagrams for computing surface and subsurface flows, respectively. Our study
 manifests an evidence that the meridional component of the flows near anti-Hale 
active regions is predominantly equatorward which disagrees with the poleward 
flow pattern seen in pro-Hale active regions. We also find clockwise or
anti-clockwise flows surrounding the anti-Joy active regions depending on their 
locations in the Southern or Northern hemispheres, respectively.
 
\end{abstract}


\maketitle

\section{Introduction}

Plasma motion in the convection zone plays a crucial role in the evolution of 
solar magnetic activity. In particular, the meridional component of the horizontal 
flow is responsible for redistributing angular momentum and transporting magnetic 
flux to the poles. It is an important ingredient in the flux-transport dynamo models
for determining the duration of the solar cycle,  the polar field strength, and  
the onset time of next solar cycle.  Therefore, precise flow  measurements are crucial 
for the better understanding of solar dynamo  \citep{Dikpati10}.  Moreover, last two solar 
minima with extended and exceptionally low magnetic activity have raised new challenges 
to our understanding of interior dynamics and the magnetic field generation \citep{Jain22a}. 
Studies suggest that the flows beneath active regions are typically larger than the quiet 
regions  \citep[e.g.,][]{Haber04,Jain15} and  in some cases  the morphology of the active 
regions alter flow directions \citep{Komm11,Irenegh13a,Jain12,Jain17}.

The emergence of active regions in each solar cycle is primarily governed by three laws;
 (i) Sp{\"o}rer’s  law postulates the latitude of emerging sunspots that decreases with time 
as the solar cycle goes forward from mid-latitudes toward the equator \citep{Sporer}, 
(ii) Hale's law governs the direction of the polarities in the regions of strong magnetic fields 
of the Sun \citep{Hale}. At any given time, the ordering of positive or negative polarities within 
the active regions with respect to the direction of rotation is same in one hemisphere, but it 
reverses in the other hemisphere. With the onset of a new solar cycle, the polarities flip 
in both hemispheres, and (iii) Joy's law defines the latitudinal dependence of the tilt of the 
bipolar active regions where leading polarity stays closer to the Equator \citep{Joy}. While 
most active regions follow these rules, there are some exceptions  and the decay of such regions 
is believed to have adverse effect on the magnetic flux transportation to the poles and ultimately 
on the strength of the polar fields. In this paper, we study flows in and around these peculiar 
active regions.

\section{Analysis and Results}

High-resolution Dopplergrams from Helioseimic and Magnetic Imager (HMI)  onboard
{\it Solar Dynamics Observatoty (SDO)} at a cadence of 45s are utilized to measure subsurface 
flows. The Dopplergrams are constructed using observations taken in the 6173.3\,\AA\, 
absorption line in the photosphere with pixel size of 0.5 arcsec. Since HMI Dopplergrams 
have some known biases, we first correct them for the systematics, e.g., the  observer motion 
and the east-west trends in Doppler velocities. We studied subsurface flows in about 180 
active regions consisting of normal and peculiar configurations, here we present the detailed 
analyses for two active regions with unusual configurations.

\subsection{Subsurface flows around anti-Hale NOAA active region 11429}

\begin{figure}
\center
\includegraphics[width=7cm]{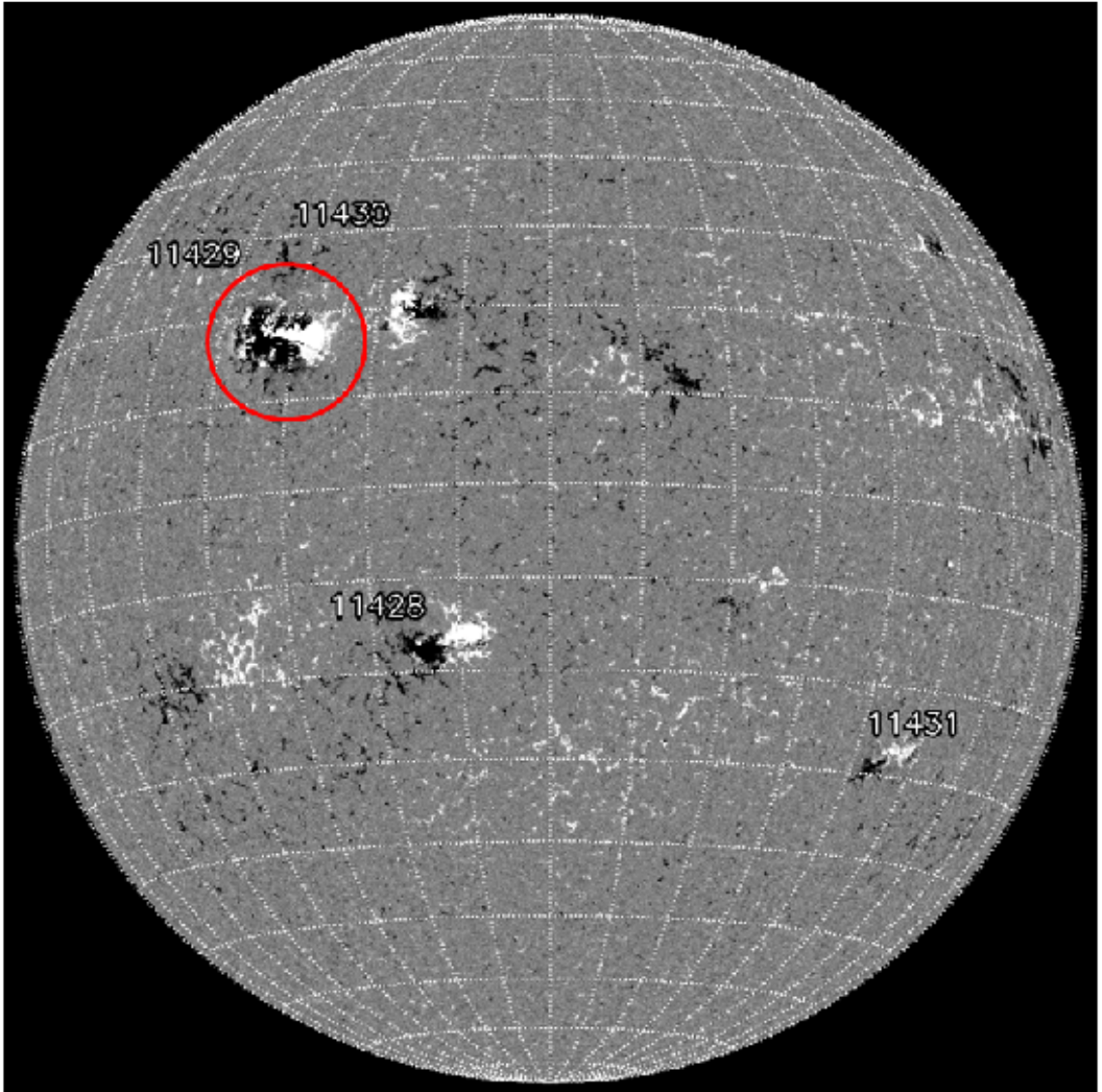}
\caption{HMI magnetogram on 2012-03-06. The anti-Hale NOAA active region 11429 is marked by a 
circle. Note that this region did violate the Hale-polarity law while other regions did not. 
Image credit: Solarmonitor.org .}
\label{fig1}
\end{figure}

NOAA 11429 was significantly large and magnetically strong active region with sunspot area 
greater than 1000 micro-Hemi and violated the Hale polarity law during its entire front-disk passage. 
It appeared on the East limb in Northern hemisphere on 2012-03-03 and crossed the West limb on 
2012-03-15. In addition, it  had a complex magnetic configuration, $\beta\gamma\delta$, and produced 
several energetic events, including X5.4-class flare. We studied flows in NOAA 11429 and compared
with those in two more regions present at the same time but with Hale-law configuration, i.e., ARs 
11430 in the close proximity of NOAA 11429 in the same hemisphere and 11428 in the opposite
 hemisphere (see Figure~\ref{fig1}). This allows us to investigate the similarities as well 
as differences in flow patterns of active regions possessing different characteristics.

For studying the subsurface flows, we use the local helioseismology technique of ring diagrams 
\citep[RD;][]{Hill88}. In this method, high-degree waves propagating in localized regions over 
the solar surface are used to obtain an average velocity vector for that region. Here, all 
three regions are tracked for 1396 minutes (i.e., 1862 HMI Dopplergrams) at the Synodgrass 
tracking rate and remapped, where each active region is centered on a 13$^\circ$ $\times$ 13$^\circ$ 
tile. A 3-D FFT is applied in both spatial and temporal directions to obtain the power spectrum 
which is fitted to a Lorentzian profile to obtain various mode parameters, including fitted 
velocities. These velocities are inverted using the regularized least-square method to obtain 
the depth dependence in both zonal and meridional components. In order to remove large scale 
flows, such as meridional circulation and rotation residuals, and the contributions from 
projection effects as discussed by \citet{Jain13}, we subtract ensemble flow averages of five 
quiet areas exactly at the same heliographic location in a nearby Carrington rotation for each 
of the individual analyzed region \citep{Tripathy09}.

\begin{figure}
\center
\includegraphics[width=13cm]{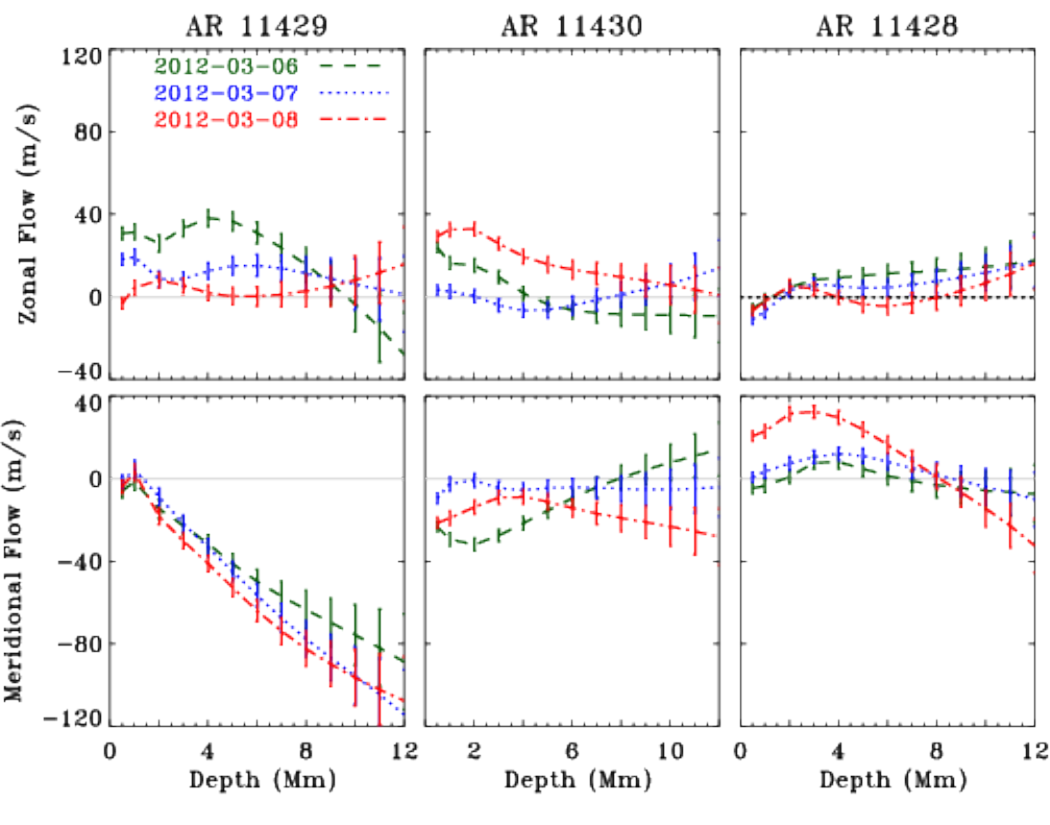}
\caption{Depth variation of zonal (top row) and meridional (bottom row) flows in 
NOAA active regions; 11429 (left), 11430 (middle) and 11428 (right) for three consecutive days. }
\label{fig2}
\end{figure}

We display, in Figure~\ref{fig2}, the zonal and meridional components of the flow vector 
for three consecutive days from March 6 -- 8. Since AR 11429 is magnetically stronger 
than other two active regions, the magnitudes of these components are also higher. The most 
significant difference can be seen in the meridional components where these are predominantly
negative (equatorward) in AR 11429 and do not change its direction with time. This component
dominates the north-south direction of the horizontal flows. It should be noted that AR 11429
was in the Northern hemisphere and in ideal scenario, it should have positive meridional flow
vector.

\begin{figure}
\center
\includegraphics[width=13cm]{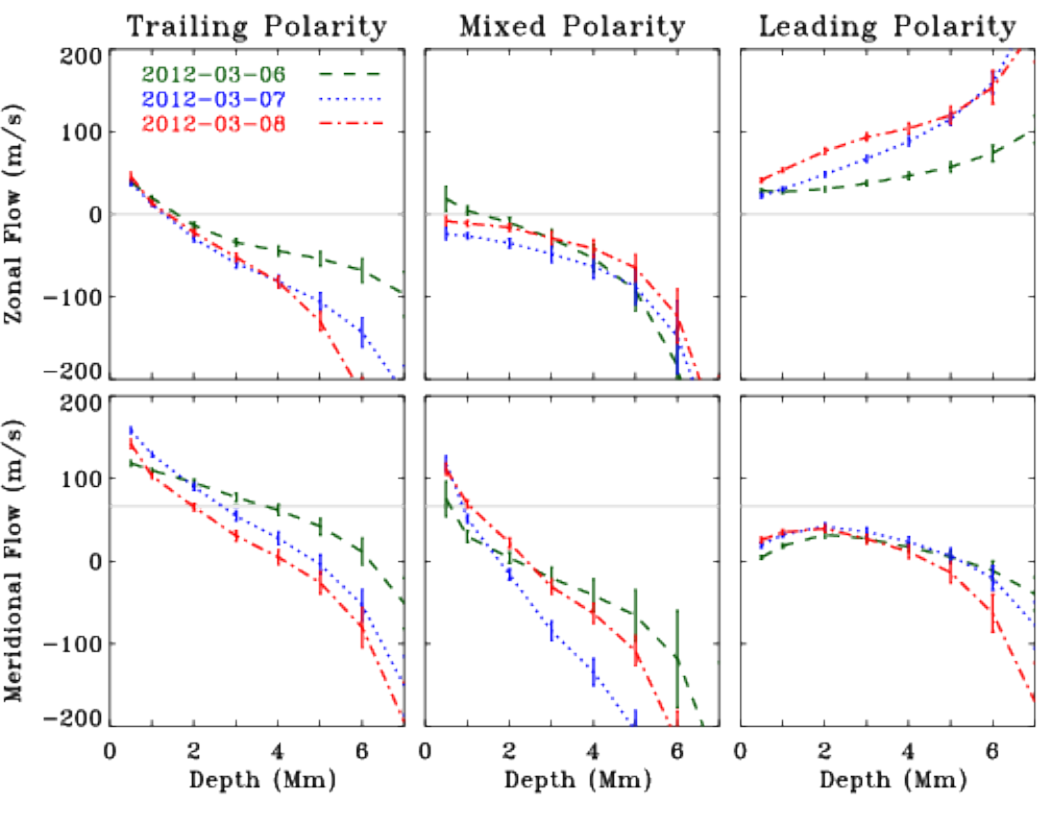}
\caption{Depth variation of zonal (top row) and meridional (bottom row) flows for three 
consecutive days for trailing-polarity (left), mixed-polarity (middle) and the leading-polarity 
(right) regions of AR 11429.}
\label{fig3}
\end{figure}

We further investigate the flow patterns in three major sub-regions within AR 11429, i.e., 
leading-polarity, mixed-polarity, and trailing-polarity regions. Here we used tiles of 
8$^o$ $\times$ 8$^o$ to minimize the effect of the nearby regions. Figure~\ref{fig3} displays 
the depth variations in zonal and meridional components for all regions. There are several 
interesting features to note in these plots: (i) there is significant variation in the velocity 
components in sub-regions as compared to flows shown in Figure~\ref{fig2} where the active region 
is considered as a whole, (ii) the zonal flows in the leading polarity region are prograde while 
these become retrograde with depth in other two regions, (iii) the meridional component of the 
leading polarity region is equatorward at all depth while it is poleward near the surface but 
becomes equatorward with increasing depth.  The horizontal flows are typically poleward in 
both hemispheres, however this characteristic changes in the case of anti-Hale active regions. 
Our study shows that the anti-Hale active region violates the normal flow pattern: the meridional 
flows in and around anti-Hale active region become predominantly equatorward that dominates 
the overall direction of the horizontal flows. These equatorward flows may have severe 
consequences on the transport of magnetic flux to the pole \citep{Jiang15}. It is argued that 
if several big regions of similar characteristics emerge during the solar cycle, these may 
affect the strength of magnetic flux at poles and finally the next cycle.

We also explore photospheric flows calculated using the Local Correlation Technique 
\citep[LCT;][]{ravindra08}. In this work, the object image is separated by 4.5 min interval 
from the reference image. The corresponding zonal and meridional velocities  are the averages 
over the entire period, i.e., 1396 minutes. We find that  zonal flows in and around anti-Hale 
active region 11429 are higher than those in two active regions. However, there are significant 
differences between the meridional flow patterns in the Northern and Southern hemispheres. 
Flows around two nearby active regions in the Northern hemisphere, i.e., anti-Hale NOAA 11429 
and pro-Hale NOAA 11430, are strongly equatorward  which is opposite to the normal flow pattern 
(poleward). Contrary to this, we do not find any unusual flow patterns in AR 11428. These flow 
measurements are consistent with those obtained with the local helioseismic technique of ring 
diagrams, as discussed in the previous subsection. Although the  spatial scales used in both 
LCT and RD methods are different, we find a strong positive correlation between the flows 
obtained from two methods near the surface which is in agreement with previous studies 
\citep[e.g.,][]{Jain16}.

\subsection{Subsurface flows around anti-Joy NOAA active region 11949}

\begin{figure}
\center
\includegraphics[width=7cm]{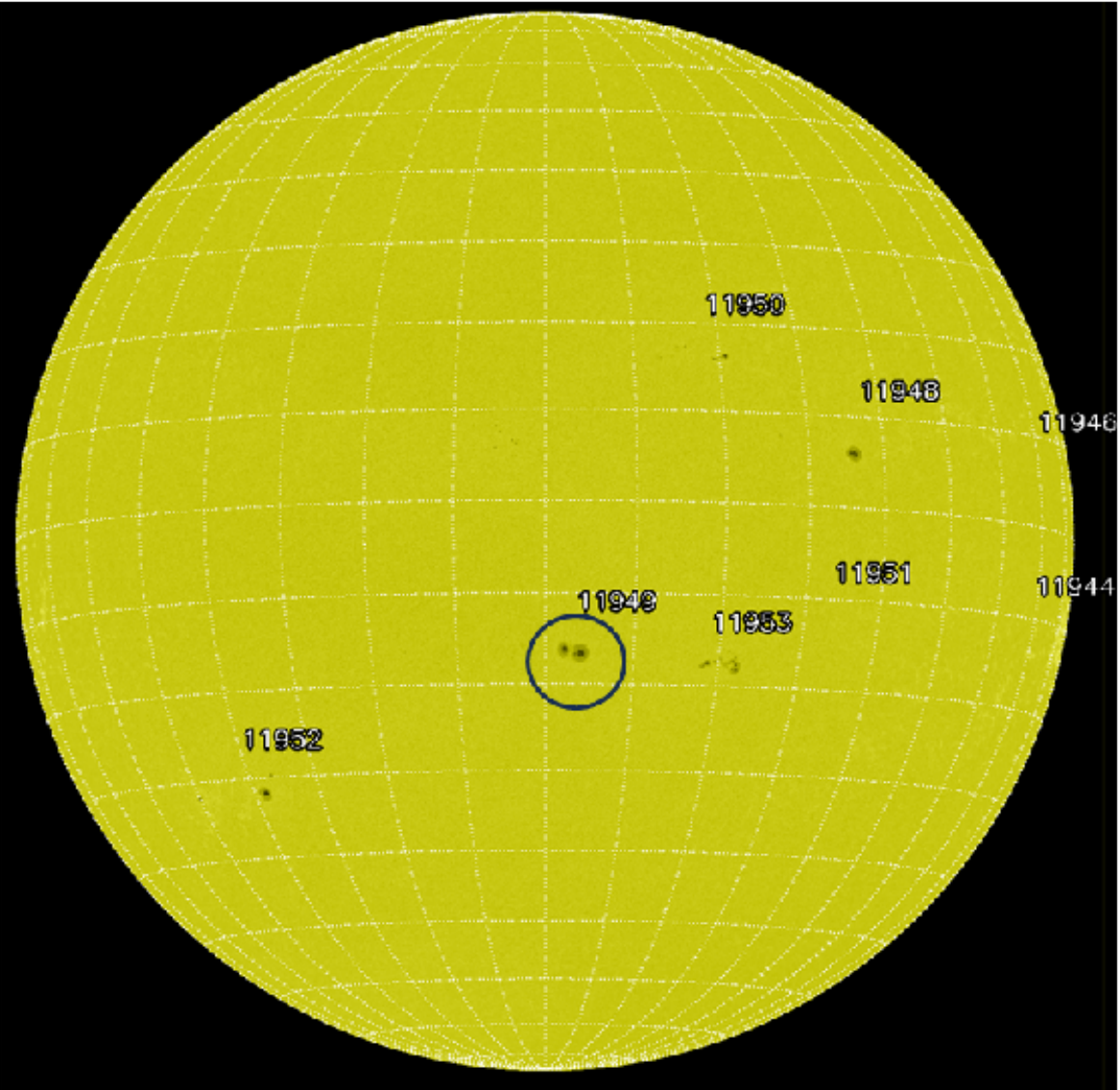}
\caption{HMI continuum image on 2014-01-14 showing the anti-Joy configuration of the
 NOAA active region 11949 (marked by a circle). It can be easily seen that the leading polarity 
lies closer to the pole than the trailing-polarity region. Image credit: Solarmonitor.org .}
\label{fig4}
\end{figure}

We selected a relatively isolated active region, i.e., NOAA 11949, that emerged in the 
Southern hemisphere around mid-January 2014. Its leading polarity is farther away from the 
equator than the trailing one. This configuration persisted for the entire disk passage 
from 2014-01-08 to 2014-01-19 (see Figure~\ref{fig4}). We studied the subsurface flows 
for five days, i.e., from 2014-01-12 to 2014-01-16, as the active region crosses the 
central meridian.

\begin{figure}
\center
\includegraphics[width=6cm]{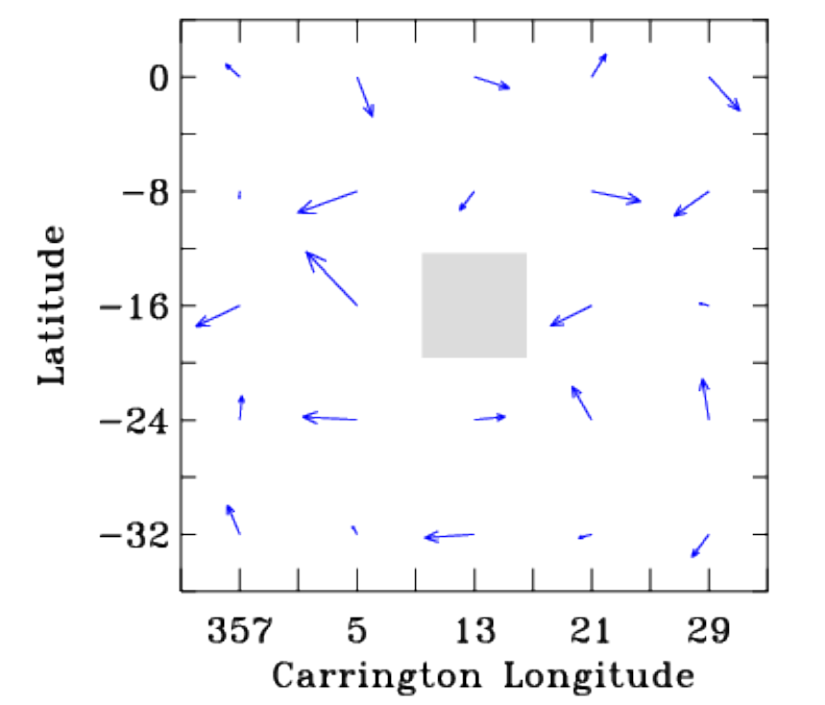}
\includegraphics[width=6cm]{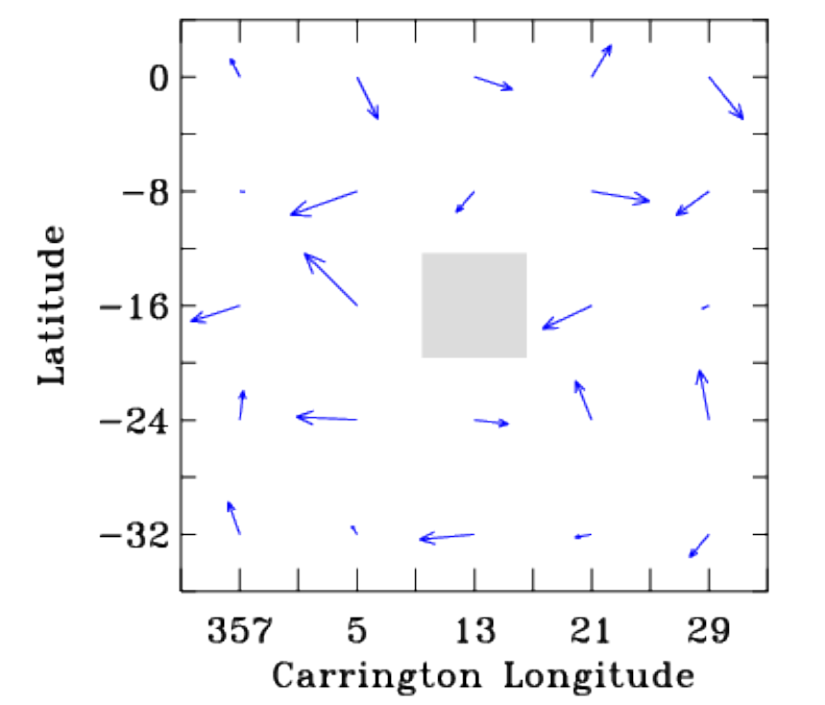}
\caption{Temporal average of the subsurface horizontal flows surrounding  NOAA active region 
11949 at 2.0 $\pm$ 1.24 Mm (left) and 7.0 $\pm$ 2.83 Mm (right) below the surface. The location 
of the region is shown by the shaded square.}
\label{fig5}
\end{figure}

To study the flows around this active region, we divided the surrounding area 
into 25 patches and followed the analysis as described for the anti-Hale active region. 
We applied ring-diagram analysis to 24 patches of 8$^\circ$ $\times$ 8$^\circ$ surrounding 
the active region and obtained the horizontal subsurface flows for all five consecutive 
days. Figure~\ref{fig5} displays temporal averages of the flows surrounding NOAA 11949 for two 
depths, 2.0 $\pm$ 1.24  Mm and 7.0 $\pm$ 2.83 Mm, below the solar surface. Since the 
subsurface properties are known to change in the presence of strong magnetic fields, here we
 concentrate on the flows around anti-Joy active region and their influence on the nearby areas.
 We obtain a curly clockwise pattern in these flows that we believe to be resulted from the 
anti-Joy configuration of NOAA 11949. Similar results were reported by \citet{Irenegh13a} 
in a case study of anti-Joy NOAA active region 11073 but counter-clockwise flows in the 
surrounding areas. We believe that the opposite flow directions in these two active regions 
are due to their locations in opposite hemispheres, which may be responsible for driving 
the leading polarity away from the solar equator. Although these two case studies exhibit 
similar results,  a statistical study comprising a large number of anti-Joy active regions 
in both hemispheres will provide a clear picture of the influence of anti-Joy regions 
on the subsurface dynamics or vice versa.

\section{Summary}
In general, the horizontal velocity in active regions are larger than the quiet regions. Zonal 
flows in anti-Hale active region do not show any significant difference from those of pro-Hale  
regions. However, we find that the leading polarity region within an anti-Hale active region 
rotates faster than the trailing polarity region.  This is opposite to the pro-Hale active 
regions. Our study provides evidence for  meridional flows around anti-Hale active regions to 
be predominantly equatorward (11 out of 14 anti-Hale active regions).  However, a thorough study 
with a larger database is required to understand the zonal and meridional flows in anti-Hale 
active regions. These equatorward flows may have severe consequences on the transport of magnetic 
flux to the pole that may affect the strength of polar flux and finally the next solar cycle. 
We also find curly flows in the surrounding areas of anti-Joy active region, which might be 
responsible for driving the leading polarity away from the solar equator.  We suggest
that the morphology of active  regions plays an important role in the subsurface flow
characteristics. 

The data used here are courtesy of NASA/SDO and the HMI Science Team. This work was supported 
by NASA grants 80NSSC19K0261, 80NSSC20K0194, 80NSSC21K0735, and 80NSSC23K0404  to the National 
Solar Observatory and the  NASA Cooperative Agreement 80NSSC22M0162 to Stanford University for the
COFFIES Drive Science Center.


\begin{thebibliography}{}
\bibitem[{{Carrington}(1858)}]{Sporer}
{Carrington}, R.~C.\ 1858, {\it MNRAS}, 19, 1

\bibitem[{{Dikpati} {et~al.}(2010){Dikpati}, {Gilman}, {de Toma}, \&
  {Ulrich}}]{Dikpati10}
{Dikpati}, M., {Gilman}, P.~A., {de Toma}, G., \& {Ulrich}, R.~K. 2010, {\it GRL}, 37, L14107

\bibitem[{{Gonz{\'a}lez Hern{\'a}ndez} {et~al.}(2013){Gonz{\'a}lez
  Hern{\'a}ndez}, {Komm}, {van Driel-Gesztelyi}, {Baker}, {Harra}, \&
  {Howe}}]{Irenegh13a}
{Gonz{\'a}lez Hern{\'a}ndez}, I., {Komm}, R., {van Driel-Gesztelyi}, L.,
  {et~al.} \ 2013, {\it JPCS}, 440,  012050

\bibitem[{{Haber} {et~al.}(2004){Haber}, {Hindman}, {Toomre}, \&
  {Thompson}}]{Haber04}
{Haber}, D.~A., {Hindman}, B.~W., {Toomre}, J., \& {Thompson}, M.~J. \  2004,
  {\it Sol Phys}, 220, 371

\bibitem[{{Hale} {et~al.}(1919){Hale}, {Ellerman}, {Nicholson}, \& {Joy}}]{Joy}
{Hale}, G.~E., {Ellerman}, F., {Nicholson}, S.~B., \& {Joy}, A.~H.\ 1919, {\it ApJ}, 49, 153

\bibitem[{{Hale} \& {Nicholson}(1925)}]{Hale}
{Hale}, G.~E., \& {Nicholson}, S.~B.\ 1925, {\it ApJ}, 62, 270

\bibitem[{{Hill}(1988)}]{Hill88}
{Hill}, F. 1988, {\it ApJ}, 333, 996

\bibitem[{{Jain} {et~al.}(2022){Jain}, {Jain}, {Tripathy}, \&  {Dikpati}}]{Jain22a}
{Jain}, K., {Jain}, N., {Tripathy}, S.~C., \& {Dikpati}, M.\ 2022, {\it ApJL}, 924, L20

\bibitem[{{Jain} {et~al.}(2012){Jain}, {Komm}, {Gonz{\'a}lez Hern{\'a}ndez},
  {Tripathy}, \& {Hill}}]{Jain12}
{Jain}, K., {Komm}, R.~W., {Gonz{\'a}lez Hern{\'a}ndez}, I., {Tripathy}, S.~C.,
  \& {Hill}, F. \ 2012, {\it Sol Phys}, 279, 349

\bibitem[{{Jain} {et~al.}(2013){Jain}, {Tripathy}, {Basu}, {Baldner}, {Bogart},
  {Hill}, \& {Howe}}]{Jain13}
{Jain}, K., {Tripathy}, S.~C., {Basu}, S., {et~al.} 2013, in Fifty Years of
  Seismology of the Sun and Stars, ed. K.~{Jain}, S.~C. {Tripathy}, F.~{Hill},
  J.~W. {Leibacher}, \& A.~A. {Pevtsov}, Vol. CS-478 (San Francisco: Astron.
  Soc. Pacific), 193

\bibitem[{{Jain} {et~al.}(2015){Jain}, {Tripathy}, \& {Hill}}]{Jain15}
{Jain}, K., {Tripathy}, S.~C., \& {Hill}, F. \ 2015, {\it ApJ}, 808, 60

\bibitem[{{Jain} {et~al.}(2017){Jain}, {Tripathy}, \& {Hill}}]{Jain17}
{Jain}, K., {Tripathy}, S.~C., \& {Hill}, F.\ 2017, {\it ApJ}, 849, 94

\bibitem[{{Jain} {et~al.}(2016){Jain}, {Tripathy}, {Ravindra}, {Komm}, \&
  {Hill}}]{Jain16}
{Jain}, K., {Tripathy}, S.~C., {Ravindra}, B., {Komm}, R., \& {Hill}, F.\ 2016,
 {\it ApJ}, 816, 5

\bibitem[{{Jiang} {et~al.}(2015){Jiang}, {Cameron}, \&
  {Sch{\"u}ssler}}]{Jiang15}
{Jiang}, J., {Cameron}, R.~H., \& {Sch{\"u}ssler}, M. 2015, {\it ApJL}, 808, L28

\bibitem[{{Komm} {et~al.}(2011){Komm}, {Howe}, {Hill}, \& {Jain}}]{Komm11}
{Komm}, R., {Howe}, R., {Hill}, F., \& {Jain}, K. \ 2011, in IAU Symposium 273 -
Physics of Sun and  Star Spots, ed. D.~P. {Choudhary} \& K.~G. {Strassmeier},
 Vol. 273,  (Cambridge: University Press), 148

\bibitem[{{Ravindra} {et~al.}(2008){Ravindra}, {Longcope}, \&
  {Abbett}}]{ravindra08}
{Ravindra}, B., {Longcope}, D.~W., \& {Abbett}, W.~P. \ 2008, {\it ApJ}, 677, 751

\bibitem[{{Tripathy} {et~al.}(2009){Tripathy}, {Antia}, {Jain}, \&
  {Hill}}]{Tripathy09}
{Tripathy}, S.~C., {Antia}, H.~M., {Jain}, K., \& {Hill}, F. \ 2009, in
  Solar-Stellar Dynamos as Revealed by Helio- and Asteroseismology: GONG
  2008/SOHO 21, ed. M.~{Dikpati}, T.~{Arentoft}, I.~{Gonz{\'a}lez
  Hern{\'a}ndez}, C.~{Lindsey}, \& F.~{Hill}, Vol. CS-416 (San Francisco:
  Astron. Soc. Pacific), 139
\end{thebibliography}
\end{document}